%% file: main.tex
\documentclass[letterpaper, 10 pt, conference]{ieeeconf}
\IEEEoverridecommandlockouts
\usepackage{amsfonts,amssymb} 
\usepackage{newtxmath}
\usepackage{algorithmic}

\input{./Texts/Initialize}
\begin{document}
\title{\huge\bf Minimax Least-Square Policy Iteration for Cost-Aware Defense of Traffic Routing against Unknown Threats}
\author{Yuzhen Zhan and Li Jin
\thanks{This work was in part supported by the National Natural Science Foundation of China, SJTU UM Joint Institute, J. Wu \& J. Sun Foundation.}
\thanks{
Y. Zhan and L. Jin are with the UM Joint Institute, Shanghai Jiao Tong University, China. (Emails: zyzhen1@sjtu.edu.cn, li.jin@sjtu.edu.cn)
}
}
\maketitle
\thispagestyle{empty}
\pagestyle{empty}

\begin{abstract}
Dynamic routing is one of the representative control scheme in transportation, production lines, and data transmission. In the modern context of connectivity and autonomy, routing decisions are potentially vulnerable to malicious attacks. In this paper, we consider the dynamic routing problem over parallel traffic links in the face of such threats. An attacker is capable of increasing or destabilizing traffic queues by strategic manipulating the nominally optimal routing decisions. A defender is capable of securing the correct routing decision. Attacking and defensive actions induce technological costs. The defender has no prior information about the attacker's strategy. We develop an least-square policy iteration algorithm for the defender to compute a cost-aware and threat-adaptive defensive strategy. The policy evaluation step computes a weight vector that minimizes the sampled temporal-difference error. We derive a concrete theoretical upper bound on the evaluation error based on the theory of value function approximation. The policy improvement step solves a minimax problem and thus iteratively computes the Markov perfect equilibrium of the security game. We also discuss the training error of the entire policy iteration process.
\end{abstract}

\input{Texts/introduction}
\input{Texts/formulation}

\input{Texts/analysis}
\input{Texts/conclusion}


\bibliographystyle{IEEEtran}
\bibliography{references}

\end{document}

%% file: Texts/Initialize.tex
\usepackage{longtable}
\usepackage{subfigure}
\usepackage{amsmath}

\usepackage{amsthm}
\allowdisplaybreaks
\usepackage{color}
\usepackage{footnote}
\usepackage{algorithm}

\usepackage{multirow} 
\usepackage{booktabs}
\usepackage{graphicx}
\usepackage{pbox}
\usepackage{breqn}
\usepackage{mathrsfs}
\usepackage{multicol}
\usepackage{supertabular}
\usepackage{enumerate}
\usepackage{url}
\usepackage[justification=centering]{caption}

\newtheorem{dfn}{Definition}
\newtheorem{thm}{Theorem}

\newtheorem{lmm}{Lemma}

%% file: Texts/introduction.tex
\section{Introduction}
Dynamic routing is a representative control problem in transportation \cite{jin2018stability}, manufacturing \cite{fraile2018trustworthy}, and networking systems \cite{laszka2019detection}.
These systems are increasingly connected and autonomous. Consequently, cyber threats (e.g., data spoofing and falsification) may mislead or tamper with nominally correct routing decisions , thus causing congestion or even destabilizing the system \cite{feng2022cybersecurity}.
Effective defensive means against such threats require either additional data for cross-validation or additional computation/storage power for encryption \cite{manshaei2013game}, both of which are costly.
In a related work \cite{xie2024cost}, the security game with complete information about the attacker was studied.
However, there are still two major challenges for designing defensive strategies, viz. (i) the lack of prior information about such threats \cite{xu2016playing} and (ii) the complex coupling between traffic state and attacker-defender game \cite{zhang2023security}.
Consequently, defensive strategies that are both cost-aware and threat-adaptive are still limited.

\begin{figure}[thpb]
      \centering
      \includegraphics[width=0.48\textwidth]{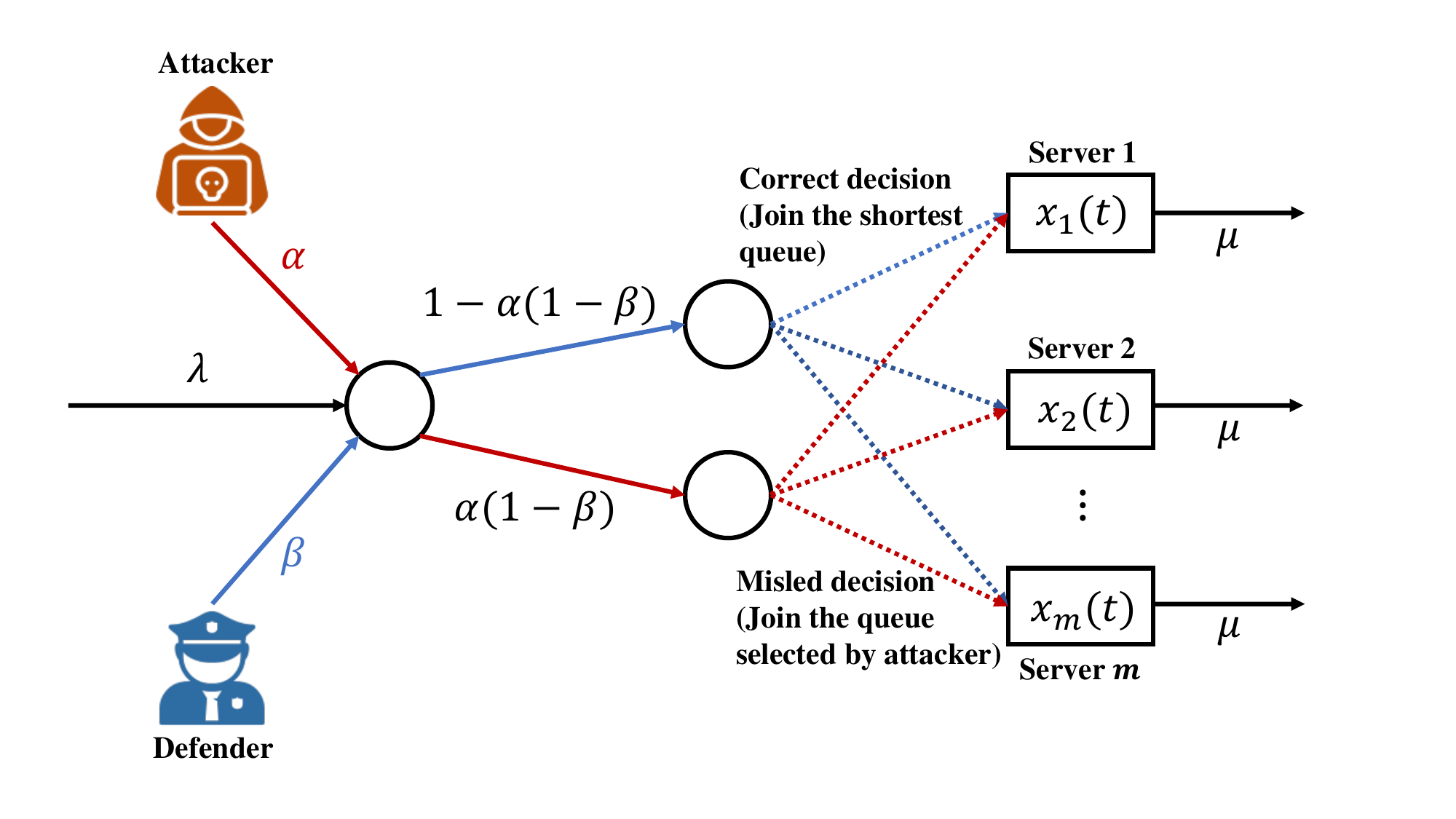}
      \caption{An $m$-queue system with shortest-queue routing under security failures.}
      \label{fig_system}
   \end{figure}

In this paper, we develop and analyze a reinforcement learning (RL) algorithm in response to the above challenge.
Our algorithm extends general policy iteration \cite{sutton2018reinforcement} to the Markov security game between an attacker and a defender (Fig.~\ref{fig_system}). The attacker can decide whether to attack or not, and the defender decides whether to protect or not.
The obtained defensive strategy is cost-aware in that it tries to balance the cost due to both traffic congestion and defensive efforts.
The strategy is also threat-adaptive in that it is responsive to the attacker's strategy, which may be unknown a priori.
In particular, we study theoretical bounds on the training error due to function approximation and stochastic sampling.

For Markov game with incomplete information, the Markov perfect equilibrium (MPE) cannot be directly derived by classical Shapley-Snow method \cite{shapley1950basic}. A natural solution is to introduce learning to the game \cite{fudenberg2016whither}. Classical algorithms in this regard include fictitious play \cite{brown1951iterative}, minimax-Q learning \cite{littman1994markov}, and policy hill-climbing \cite{bowling2002multiagent}. Recently, there is an increasing interest in RL for Markov games. Zhang et al. \cite{zhang2020model} demonstrated the sample-complexity of model-based approach to Nash equilibrium. Pérolat et al. \cite{perolat2016softened} made use of quasi-Newton methods to minimize different norms of the optimal Bellman residual of Markov game. Mao et al. \cite{mao2023provably} proposed a variant of Q-learning and proves convergence to an approximated error. Sayin et al. \cite{sayin2021decentralized} develops a radically uncoupled Q-learning dynamics for zero-sum Markov games. The above results provide useful hints for our problem. However, we still lack explicit error bounds for RL with value function approximation in such Markov games.
 


%
In this paper, we build on the theory of least-square policy iteration (LSPI \cite{lagoudakis2003least}) and propose a solution algorithm for the Markov security game (Fig.~\ref{fig_system}) with theoretical guarantees. We consider a queuing system with Poisson arrivals and homogeneous parallel servers. Without attacks, the nominally optimal routing policy, the ``join-the-shortest-queue'' policy, is applied. An attacker (resp. defender) is able to manipulate (resp. secure) the correct routing decision, at a technological cost, to increase (resp. decrease) traffic queues. We consider a zero-sum game between these two players; more details about the modeling and formulation are available in \cite{xie2024cost}.
We suppose that the defender has no prior knowledge of the attacker's strategy.
Since the queuing system has a large state space, we consider a linear approximation of the value/cost function \cite{sutton2018reinforcement}. To compute the MPE, we extend the classical LSPI for Markov decision processes \cite{lagoudakis2003least} to the game-theoretic setting by incorporating a minimax problem in policy improvement.

Our main result (Theorem 1) is a theoretical bound on the prediction error of the proposed algorithm. We establish the bound by decomposing the prediction error into the projection error and the sampling error. The projection error results from the structure of the approximate value function, and is typically empirically estimated. To bound the sampling error, we utilize model of regression with Markov design \cite{lazaric2012finite} and further decompose it into the sampling error for the approximate value function and that for the true value function. The former is estimated by applying regression bound of Markov design. The latter is estimated by studying the concentration property of reward function and other properties of the queuing systems, with concentration properties of the transition dynamics established \cite{osband2017posterior}.

Our main contributions include:
\begin{enumerate}
  \item An evaluation algorithm that computes the value for any policy pair, requiring no knowledge/observation of the attacker's policy;
  \item A theoretical bound on the evaluation error, explicitly in terms of model parameters; 
  \item A policy iteration algorithm that estimates the MPE. 
\end{enumerate}


The rest of this paper is organized as follows. Section~\ref{sec_model} formulates the queuing model, the Markov game, and the approximate value function. Section~\ref{sec_pe} presents the evaluation scheme and derives the bound for evaluation error. Section~\ref{sec_pl} gives the minimax LSPI algorithm and discusses its convergence.
Section~\ref{sec_con} gives the concluding remarks.

%% file: Texts/formulation.tex
\section{Model formulation}
\label{sec_model}

In this section, we introduce the queuing system model, formulate the Markov security game, and develop a linear approximation for the action value function of the game.

\subsection{System and player models}
\label{sub_qs}

Consider the parallel queues in Fig.~\ref{fig_system}. Jobs arrive according to a Poisson process of rate $\lambda>0$ and go to one out of $m$ identical serves, each with exponentially distributed service times and with service rate $\mu>0$. Let $x(t)\in S:=\{0,1,\ldots,L\}^m$ be the vector of the number of jobs in the servers, either waiting or being served; $L$ is the buffer size. In the absence of attacks, an incoming job is routed to the server with the shortest queue; ties are broken uniformly at random. If a job is routed to a fully occupied server, it is rejected and never returns.

An attacker is able to manipulate the routing decision for an incoming job. The attacking cost is $c_a>0$ per unit time. 
A defender is able to defend the routing decision for an incoming job, at a cost of $c_b>0$ per unit time.
If a routing decision is attacked and is not defended, the job will go to a server selected by the attacker. Otherwise, the job will join the shortest queue. We refer the readers to \cite{xie2024cost} for details on the justification and interpretation of the player model.

The action space for the attacker is $A=\{0,1\}$, where $a(t)=0$ (resp. $a(t)=1$) means ``not attack'' (resp. ``attack''.) The actions space $B=\{0,1\}$ and action $b(t)$ for the defender is defined analogously.
Then, the instantaneous reward (resp. cost) for the attacker (resp. defender) at time $t$ is defined as
\begin{align}
    \rho(x(t),b(t),a(t)):= \Vert x(t) \Vert_{1}-c_aa(t)+ c_bb(t),
    \label{Reward}
\end{align}
where $\|\cdot\|_1$ is the 1-norm. The technology costs terms are motivated by the agent potential interest in maximizing opponent's costs. 

\subsection{Markov security game}
\label{sub_mg}
Since the system state is countable, we can formulate the Markov security game between the attacker and the defender in discrete time (DT). Specifically, let $t_k$ be the $k$th transition epoch of the continuous-time process. With a slight abuse of notation, let
\begin{align*}
    x_k=x(t_k),\ 
    a_k=a(t_k),\ 
    b_k=b(t_k),\quad k=0,1,...
\end{align*}
Thus, the transition probabilities $p(x'|x,a,b)$ for the DT process can be obtained by the theory of countable-state Markov processes \cite{gallager2013stochastic}.
In addition, the expected one-step reward for the defender is given by
\begin{align}
    r\Big(x_k,a_k,b_k\Big):=\rho\Big(x(t_{k-1}),a(t_{k-1}),b(t_{k-1})\Big)\mathbb E[\Delta t_k],
     \label{reward}
\end{align}
where $\Delta t_k=t_k-t_{k-1}$ is the exponentially distributed inter-transition interval.

We use $\alpha(a|x)$ to denote the strategy for the attacker and $\beta(b|x)$ for the defender. The attacker (resp. defender) attempts to maximize (resp. minimize) the expected cumulative discounted reward (resp. cost) given by
$$
v_{\alpha,\beta}(x)=\mathbb{E}_{\alpha,\beta}\Big[\sum\limits_{k}^{\infty}{\gamma^k r(x_k,a_k,b_k)\Big|x_0=x}\Big].
$$
The corresponding action value is given by
$$
q_{\alpha,\beta}(x,a,b)
=r(x,a,b)+\sum_{x'\in S}p(x'|x,a,b)v_{\alpha,\beta}(x').
$$



\begin{dfn}
    The Markov perfect equilibrium (MPE) for the security game is a strategy pair $(\alpha^*,\beta^*)$ such that for any $x \in \mathcal{S}$,
\begin{align*}
   &\alpha^*(x) = \arg \max_{\alpha}v_{\alpha,\beta^*}(x),\\
   &\beta^*(x) = \arg \min_{\beta}v_{\alpha^*,\beta}(x).
\end{align*}
\end{dfn}


Hence, the MPE is characterized by the equilibrium state value function
$$
v^*(x)=v_{\alpha^*,\beta^*}(x).
$$
By the Shapley theory \cite{shapley1953stochastic}, $v^*$ is associated with a unique action value function (also called the ``minimax $Q$ function'') satisfying the minimax version of the Bellman optimality equation
\begin{align*}
    &q^*(x,a,b)= r(x,a,b)\nonumber\\
    &\quad+\gamma\min_{\beta} \max_{a'}\sum_{\substack{x'\in S\\b'\in B}}p(x'|x,a',b)\beta(b'|x)q^*(x',a',b')\\
    &= r(x,a,b)\nonumber\\
    &\quad+\gamma\max_{\alpha} \min_{b'}\sum_{\substack{x'\in S\\a'\in A}}p(x'|x,a',b)\alpha(a'|x)q^*(x',a',b').\nonumber
\end{align*}
Following \cite{zhu2020online}, we take the defender's perspective also write $q^*$ as the fixed point for the minimax Bellman operator $T_\beta$ given by
\begin{align*}
    &(T_\beta q^*)(x,a,b)= r(x,a,b)\nonumber\\
    &+\gamma\min_{\beta} \max_{a'}\sum_{\substack{x'\in S\\b'\in B}}p(x'|x,a',b)\beta(b'|x)q^*(x',a',b');
\end{align*}
Note that there is a symmetric operator for the attacker.

\subsection{Function approximation}
\label{sub_fa}
Since analytical solution for the value function is difficult, we consider a linear approximation in the following form:
\begin{align*}
    \hat{q}(x,a,b;\theta) = \sum_{i=1}^{d} \mathnormal{\phi}_i(x,a,b)\theta_{i},
\end{align*}
where $d=m+2$ is the number of feature functions, $\theta = [\theta^1, \ldots, \theta^d] \in \mathbb{R}^d$ is weight vector; the feature functions are constructed, based on insights about the system dynamics, by
\begin{align}
\phi_i(x,a,b)=
    \begin{cases}
(x_i+\delta_i(a,b))^2 & \mbox{if }i\in [1,\ldots,d-2],\\
 a & \mbox{if }i=d-1,\\
 b & \mbox{if }i=d,
 \label{feature_func}
    \end{cases}
\end{align}
where $\delta_i(a,b)$ is given by
\begin{align*}
\delta_i(a,b)=
    \begin{cases}
 1 & \mbox{if }i = \arg\max_{i} x_i,(a,b)=(1,0),\\
 1 & \mbox{if }i=  \arg\min_{i} x_i,(a,b)\neq (1,0), \\
 0 & \mbox{otherwise}.
    \end{cases}
\end{align*}
Intuitively, $\phi_1,\phi_2,\ldots,\phi_{d-2}$ are motivated by the first term in the right-hand side of \eqref{Reward}, while $\phi_{d-1},\phi_{d}$ are associated with the other two terms, respectively.
Note that $\phi_i$ are linearly independent.



Suppose a sample set $\mathcal{D}_{\alpha,\beta} = \{(x_k,a_k,b_k, r_k);k=1,2,\ldots,n\}$ generated by a pair of stationary policies $(\alpha,\beta)$ for both players. Let
\begin{align*}
Q_{\alpha,\beta}=\left[\begin{matrix}
    q(x_1,a_1,b_1) \\
    q(x_2,a_2,b_2) \\
    \vdots \\
    q(x_n,a_n,b_n) \\
\end{matrix}\right],\quad
\Phi_{\alpha,\beta}=\left[\begin{matrix}
    \mathnormal{\phi}(x_1,a_1,b_1) \\
    \mathnormal{\phi}(x_2,a_2,b_2) \\
    \vdots \\
    \mathnormal{\phi}(x_n,a_n,b_n)
\end{matrix}\right].
\end{align*}%
Let $\mathscr{Q}_{\alpha,\beta}=\{ \Phi_{\alpha,\beta}\theta, \theta \in \mathbb{R}^d \}$ be the space of ${\hat Q}_{\alpha,\beta}(\theta)$.
We define the $\sigma$-norm for any $y \in \mathbb{R}^n$ as
\begin{align*}
    \Vert y \Vert_{\sigma}^{2} = \frac{1}{n}\Vert y \Vert_{2}^{2}.
\end{align*}
Then the orthogonal projection of $Q_{\alpha,\beta}$ onto $\mathscr{Q}_{\alpha,\beta}$ is given by
\begin{align*}
    \hat{\Pi}_{\alpha,\beta}Q_{\alpha,\beta} = \arg \min_{\hat{Q}\in \mathscr{Q}_{\alpha,\beta}} \Vert Q_{\alpha,\beta}-\hat{Q} \Vert_{\sigma},
\end{align*}
where 
$$
\hat{\Pi}_{\alpha,\beta}
=\Phi_{\alpha,\beta}(\Phi^{\top}_{\alpha,\beta}\Phi_{\alpha,\beta})^{-1}\Phi^{\top}_{\alpha,\beta}
$$
is useful for policy evaluation \cite{lazaric2010finite}.

The empirical transition probability in ${\mathcal D}_{\alpha,\beta}$ is given by
$$
\hat{p}(x^{\prime}|x,a,b)=\begin{cases}
    \frac{\sum_{k=1}^{n-1}\mathbb I\{x_k=x,a_k=a,b_k=b,x_{k+1}=x'\}}{\sum_{k=1}^{n-1}\mathbb I\{x_k=x,a_k=a,b_k=b\}}\\
    \hspace{3cm}(x,a,b,\cdot)\in\mathcal D,\\
    0\hspace{2.8cm} \text{otherwise.}
\end{cases}
$$
Also define
$$
w(x,a,b)=\begin{cases}
    \Big(\sum_{k=1}^{n}\mathbb I\{x_k=x,a_k=a,b_k=b\}\Big)^{-\frac12},\\
    \hspace{3cm}(x,a,b,\cdot)\in\mathcal D,\\
    \frac{mL+c_b}{\lambda(1-\gamma)}\hspace{2.1cm} \text{otherwise.}
\end{cases}
$$

\noindent{\bf Assumption 1}. \textit{Given a sample set $\mathcal{D}_{\alpha,\beta}$, there exists a constant $C_P$ such that $\forall (x_k,a_k,b_k,r_k) \in \mathcal{D}$, the following relationships hold with high probability $1-\delta$}
\begin{align*}
   &\vert \hat{p}(x^\prime|x_k,a_k,b_k)- p(x^\prime|x_k,a_k,b_k) \vert\\
   &\hspace{2cm}\leq C_P \sqrt{\log(1 / \delta)}w(x_k,a_k,b_k),
\end{align*}

The above assumption essentially constraints the ``continuity'' of $\hat p$ \cite{kumar2020conservative}.
Note that the constant $C_P$ depends on the difference between the empirical transition matrix and the true one, induced by the sample distribution \cite{osband2017posterior,auer2008near}.
We also write
\begin{align}
\label{hatP}
\setlength{\arraycolsep}{3pt}
    \hat{P}=\left[\begin{matrix}
    \hat{p}(x_{1}|x_1,a_1,b_1) & \ldots      &\hat{p}(x_n|x_1,a_1,b_1)\\
    \vdots               & \ddots      &\vdots                 \\
    \hat{p}(x_{1}|x_{n},a_{n},b_{n}) & \ldots      &\hat{p}(x_{n}|x_{n},a_{n},b_{n})
\end{matrix}\right],
\end{align}
and
$
W=[
    w(x_1,a_1,b_1),\cdots,
    w(x_{n},a_{n},b_{n})]^T.
$
Note that all the above matrices/vectors are derived from the data set $\mathcal D_{\alpha,\beta}$.

%% file: Texts/analysis.tex
\section{Policy Evaluation}
\label{sec_pe}
In this section, we develop and study a policy evaluation algorithm for the security game.
Since the policies to be evaluated are fixed in this step, for ease of presentation, we drop the subscripts of $\alpha,\beta$ in this section; i.e.,
\begin{align*}
    Q=Q_{\alpha,\beta},\ 
    \Phi=\Phi_{\alpha,\beta},\ 
    \hat Q={\hat Q}_\beta,\ 
    T=T_\beta,\ 
    {\hat\Pi}={\hat\Pi}_{\alpha,\beta}.
\end{align*}
Given a pair of policies $(\alpha,\beta)$ and $\mathcal{D}_{\alpha,\beta}$, the evaluation task is to find $\theta^*$ that solves
\begin{align*}
    \min_{\theta\in{\mathbb R}^d} \Big\|\hat{Q}(\theta)-T\hat{Q}(\theta)\Big\|_\sigma.
\end{align*}
The optimal solution turns out to be
\begin{equation}
\label{thetaup}
    \begin{aligned}
    \theta^*&= \left (\Phi^{\top}(\Phi-\gamma P \mathrm{B} \max_{a}\Phi) \right )^{-1} \Phi^{\top}R,
\end{aligned}
\end{equation}
where
\begin{align*}
\setlength{\arraycolsep}{3pt}
    P=\left[\begin{matrix}
    p(x_{1}|x_1,a_1,b_1) & \ldots      &p(x_{n}|x_1,a_1,b_1)\\
    \vdots               & \ddots      &\vdots                 \\
    p(x_{1}|x_{n},a_{n},b_{n}) & \ldots      &p(x_{n}|x_{n},a_{n},b_{n})
\end{matrix}\right],
\end{align*}
\begin{align*}
\mathrm{B}=\left[\begin{matrix}
    \beta(b_1 | x_1) & \ldots  & \beta(b_n|x_1)\\
    \vdots         & \ddots  &\vdots         \\
    \beta(b_1 | x_{n}) & \ldots & \beta(b_n | x_{n})
\end{matrix}\right], \quad
R=\left[\begin{matrix}
    r(x_1,a_1,b_1)  \\
    \vdots \\
    r(x_n,a_n,b_n) \\
\end{matrix}\right].
\end{align*}
Then, $\hat Q=\hat{Q}(\theta^*)$ is the estimate of the action value function.

The main result of this paper is a theoretical bound on the evaluation error:



\begin{thm}
Consider the security game on a parallel service system with $m$ servers of buffer size $L$. Let $\lambda$ be the arrival rate, $\gamma$ be the discount rate, $c_b$ the the defending cost, respectively. Let $C_p$ be the constant and $W$ be the vector defined in Assumption 1.
Suppose a sample set $\mathcal{D}$ of size $n$ generated under a policy pair $(\alpha,\beta)$ for the game. Then, with probability at least $1-\delta$,
    \begin{align*}
        &\Big\Vert Q-\hat{Q}  \Big\Vert_{\sigma} \nonumber\\
        &
        \le\frac{1}{\sqrt{1-\gamma^2}}\Big\Vert Q-\hat{\Pi}Q\Big\Vert_{\sigma}+\frac{\gamma L^2 (mL+c_b)}{\lambda(1-\gamma)^3}\sqrt{\frac{2(m+2)\log (2/ \delta)}{nv_{min}}} \nonumber\\
        &  \quad+
        \frac{1}{\sqrt{n}} \left \Vert \hat{\Pi}\frac{mL+c_b}{\lambda}\left (\mathbf{1}_n+\frac{\gamma}{1-\gamma}C_P \sqrt{\log(1 / \delta)}W \right) \right \Vert_2,
    \end{align*}
    where $\nu_{\min}$ is the smallest eigenvalue of $\frac{1}{n}\Phi^\top\Phi$.

\end{thm}

The three terms on the right-hand side in the above correspond to the projection error $e_p$, the sampling error for true value function $e_{st}$, and the sampling error for the approximate value function $e_{sa}$. One can see that the sampling errors vanish as $n\to\infty$. The rest of this section is devoted to the proof of the theorem.

\subsection{Decomposition}
\label{error_1}
By (\ref{thetaup}), $\hat{Q}=\Phi \theta^*$. To get an unbised approximation of $\theta^*$, the policy evaluation step should use Bellman operator.  However, $\mathcal{D}_{\alpha,\beta}$ typically does not contain all possible transitions, this step actually uses an empirical Bellman operator defined by
\begin{align*}
     [\hat{T}q](x,a,b) = \hat{r}(x,a,b)+\gamma\mathbb{E}_{\hat{P}}\max_{a'\in A}\sum_{b'}\beta(b'|x')q(x',a',b'),
\end{align*}
where $\hat{r}(x,a,b)$ is given by $\mathcal{D}_{\alpha,\beta}$. $\hat{P}$ is the empirical transition matrix defined in (\ref{hatP}).
Thus there exists a sampling error on $\hat{Q}$.
In the following sections, we denote the $\hat{Q}$ with sampling error as $\hat{Q}_s$.



We first decompose the error w.r.t $\Vert Q-\hat{Q}_s \Vert_{\sigma}$ and derive its bound. Then we analyze the bound of $\Vert \hat{Q}-\hat{Q}_s \Vert_{\sigma}$. Finally by the triangle inequality we can derive the one of $ \Vert Q-\hat{Q} \Vert_{\sigma}$.

We aim to show the existence and uniqueness of the $\hat{Q}$ under the operator $\hat{\Pi}\hat{T}$. The proof is adapted to a TD learning setting with function approximation from the Theorem 3.9 in \cite{van1998learning}. The fundamental property employed to demonstrate the convergence and finite sample bound of on-policy TD learning with linear function approximation, as outlined in the influential study \cite{tsitsiklis1996analysis}, is that the associated Bellman operator is a contraction mapping not solely w.r.t the $l_\infty$ norm, but also w.r.t to a weighted $l_2$ norm. And our definition of the $\sigma$ norm applies a uniform weight on the $l_2$ norm.

\begin{lmm}
Let $\hat{\Pi}$ be the orthogonal projection operator that projects onto the space of $\mathscr{Q}_n$. Then, the combination $\hat{\Pi}\hat{T}$ is a contraction with contraction factor $\kappa \leq \gamma$, and has a unique fixed point $\hat{Q}_s$.
\end{lmm}
\noindent\textit{Proof.}
For $Q \in \mathbb{R}^n$, we obtain
\begin{align*}
   \Vert \hat{\Pi}Q \Vert_{\sigma}^2=\langle Q,\hat{\Pi}Q \rangle_{\sigma}\leq
   \Vert Q \Vert_{\sigma} \Vert \hat{\Pi}Q \Vert_{\sigma},
\end{align*}
by the Cauchy-Schwarz inequality we have $\Vert \hat{\Pi}Q \Vert_{\sigma} \leq \Vert Q \Vert_{\sigma}$. Thus $\hat{\Pi}$ is non-expansive. Then for any $Q_1, Q_2 \in \mathbb{R}^n$ we have
\begin{align*}
   \Vert \hat{\Pi}\hat{T} Q_1 - \hat{\Pi}\hat{T}Q_2 \Vert_{\sigma} \leq 
   \Vert \hat{T}Q_1 - \hat{T}Q_2  \Vert_{\sigma} \leq \gamma \Vert Q_1 - Q_2  \Vert_{\sigma}.
\end{align*}
Thus $\hat{\Pi}\hat{T}$ is a contraction with a contraction factor $\kappa \leq \gamma$. By applying the Banach fixed point theorem, there exists a unique fixed point of $\hat{\Pi}\hat{T}$ such that $\hat{Q}_s=\hat{\Pi}\hat{T}\hat{Q}_s$.

$\qquad \qquad \qquad \qquad \qquad \qquad \qquad \qquad \qquad \qquad  \qquad \ \ \ \qedsymbol$

By the Pythagorean Theorem and triangle inequality
\begin{align*}
    \Vert Q-\hat{Q}_s \Vert_{\sigma}^2 &= \Vert Q-\hat{\Pi}Q \Vert_{\sigma}^2+\Vert \hat{Q}_s-\hat{\Pi}Q \Vert_{\sigma}^2 \\& \leq \Vert Q-\hat{\Pi}Q \Vert_{\sigma}^2 \\& \quad + \ ( \Vert \hat{Q}_s-\hat{\Pi}\hat{T}Q \Vert_{\sigma} + \Vert \hat{\Pi}\hat{T}Q - \hat{\Pi}Q \Vert_{\sigma})^2.
\end{align*}
From the contraction fixed-point results of Lemma 1, we have 
\begin{align*}
    \Vert \hat{Q}_s-\hat{\Pi}\hat{T}Q \Vert_{\sigma} = \Vert \hat{\Pi}\hat{T}\hat{Q}_s-\hat{\Pi}\hat{T}Q \Vert_{\sigma} \leq \gamma \Vert Q-\hat{Q}_s \Vert_{\sigma} .
\end{align*}
It follows that 
\begin{equation}
\begin{aligned}
\label{decomposition}
    \Vert Q-\hat{Q}_s \Vert_{\sigma}^2 &\leq \Vert Q-\hat{\Pi}Q \Vert_{\sigma}^2\\& \quad + (\gamma \Vert Q-\hat{Q}_s \Vert_{\sigma}+ \Vert \hat{\Pi}\hat{T}Q - \hat{\Pi}Q \Vert_{\sigma})^2.
\end{aligned}
\end{equation}

\begin{figure}[H] 
\centering 
\includegraphics[width=0.4\textwidth]{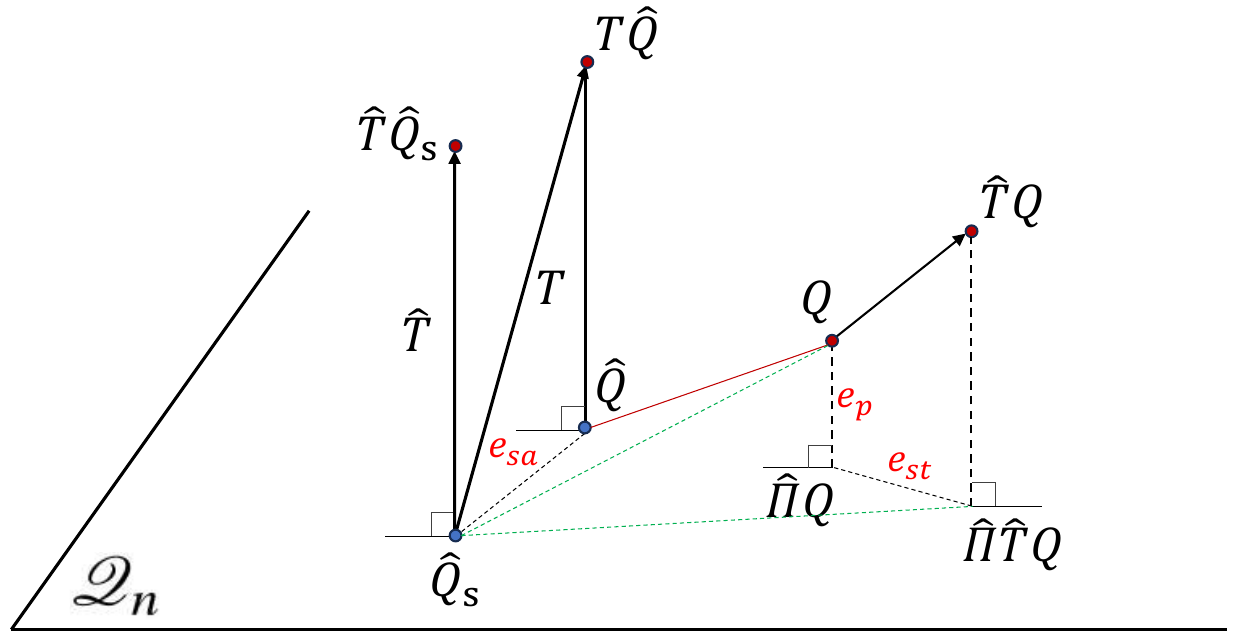} 
\caption{Components used in proof of Theorem 1.} 
\label{fig_2} 
\end{figure}

In Figure \ref{fig_2}. Red line is what we're asking for. Green dashed lines are mediators, which can be transformed by known inequalities or contraction properties. Red words indicate distances in the bound of Theorem 1,among which $e_{p}$ denotes projection error, $e_{st}$ and $e_{sa}$ denote sampling error respectively for true value function and approximate value function.

\subsection{Sampling error for true value function}
\label{error_2}
This subsection is devoted to the analysis of bound of term $\Vert \hat{\Pi}\hat{T}Q - \hat{\Pi}Q \Vert_{\sigma}$ in decomposition equation (\ref{decomposition}).

\begin{dfn}
    Given $\mathcal{D}_{\alpha,\beta}$, $(X_k)_{1 \leq k \leq n}=(x_k,a_k,b_k)_{1 \leq k \leq n}$, $(\tilde{Z}_k)_{1 \leq k \leq n}$ are generated as follows:  Define $f$ as the target function of the policy evaluation step, then $\tilde{Z}_k=f(x_k,a_k,b_k)+\zeta_k$, where $\zeta$ denotes a random noisy variable such that
\begin{align*}
    \vert \zeta_k \vert \leq C \quad and \quad \mathbb{E}[\zeta_k|X_1,\ldots,X_k]=0.
\end{align*}
\end{dfn}



The following lemma is essentially built on the model of regression with Markov design. We gain insight from \cite{lazaric2012finite} and implement a similar analysis. The main difference lies in that we take tuple input as generalization and delete the union bound over all features, thus shrink the bound by $\mathcal{O}(\sqrt{d)}$, where $d$ is the number of feature functions.

\begin{figure}[H] 
\centering 
\includegraphics[width=0.4\textwidth]{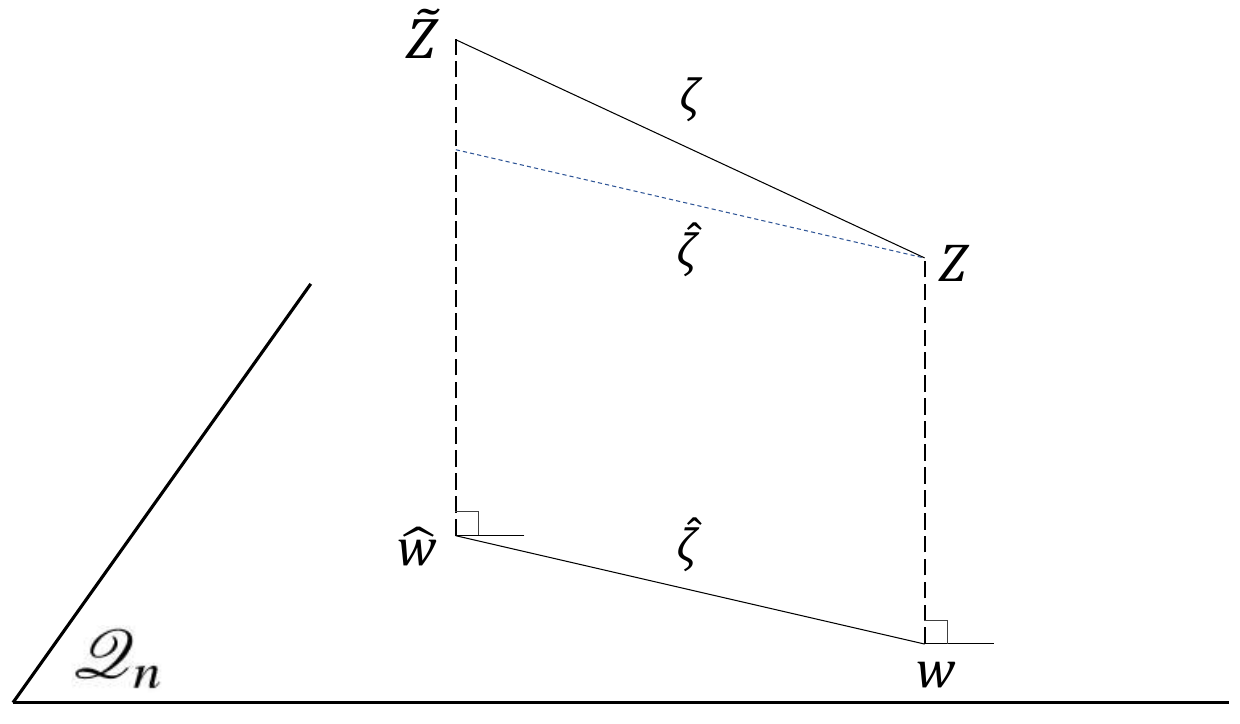} 
\caption{Components used in proof of Lemma 2} 
\label{fig_1} 
\end{figure}

\begin{lmm}
    Let $w \in \mathscr{Q}_n$ be the least-squares estimate of vector $Z=\{Z_k=f(X_k)\}_{k=1}^n$,i.e. $w=\hat{\Pi}Z$. Let $\hat{w} \in \mathscr{Q}_n$ be the least-squares estimate of vector $\tilde{Z}=\{\tilde{Z}_k\}_{k=1}^n$, i.e. $\hat{w}=\hat{\Pi}\tilde{Z}$. Then for any $\delta$ w.r.t to the random $(X_k)_{1 \leq k \leq n}$, with probability at least $1-\delta$,
    \begin{align*}
        \Vert \hat{w}-w \Vert_{\sigma} \leq CL^2\sqrt{\frac{2d\log (2/ \delta)}{nv_{min}}}.
    \end{align*}
\end{lmm}



\noindent\textit{Proof.}  
Define $\zeta \in \mathbb{R}^n$ as the vector with components $\zeta_k = \tilde{Z}_k - Z_k$. Let $\hat{\zeta}$ be the projection of $\zeta$ to the space $\mathscr{Q}_n$, that is, $\hat{\zeta}=\hat{\Pi}\zeta=\hat{\Pi}(\tilde{Z} - Z)=\hat{w}-w$. Thus the associated parameter of the projection is denoted by $\theta \in \mathbb{R}^d$ such that $\hat{\zeta}=\Phi\theta$. Note that the parameter $\theta$ may not be unique when the sample-based Gram matrix $\frac{1}{n}\Phi^\top\Phi$ has zero eigenvalue, then $\Phi^\top\Phi$ is not invertible. Thus $\hat{\Pi}=\Phi(\Phi^\top\Phi)^{-1}\Phi^\top$ can not be well-defined and $\theta$ may have many solutions. We choose the one with minimal $l_2$-norm, i.e. $\theta=\Phi^{+}\hat{\zeta}$ in which case its component in null-space of $\Phi$ will be zero. Since the projection is orthogonal, we have $\langle\hat{\zeta},\zeta\rangle_{\sigma}=\Vert \hat{\zeta} \Vert_{\sigma}^{2}$. Then by Cauchy-Schwarz inequality we can deduce
\begin{equation}
\begin{aligned}
\label{lemma_markovdesign_1}
   \Vert \hat{\zeta} \Vert_{\sigma}^{2}&=\langle\hat{\zeta},\zeta\rangle_{\sigma}=\frac{1}{n}\sum_{i=1}^{d}\theta^{i}\sum_{k=1}^{n}\zeta_k\Phi_{i}(X_k)\\ &\leq \frac{1}{n} \Vert \theta \Vert_2 \left [\sum_{i=1}^{d}(\sum_{k=1}^{n}\zeta_k\Phi_{i}(X_k))^2 \right ]^{\frac{1}{2}}.
\end{aligned}
\end{equation}

The null-space of $\Phi$ is denoted by $K$, thus $\theta$ can be decomposed with component $\theta(K) \in K$ and $\theta({K^{\perp}}) \in K^{\perp}$. Since the decomposition is also orthogonal, it can be referred that $\Vert \theta \Vert_{2}^{2}=\Vert \theta(K) \Vert_{2}^{2} + \Vert \theta({K^{\perp}}) \Vert_{2}^{2}$. 

It is known that the Gram matrix $\frac{1}{n}\Phi^\top\Phi$ is positive-semidefinite. Then the null-space $K$ holds zero eigenvalues and $K^{\perp}$ holds others. From the assumption that the smallest strictly positive eigenvalue of $\frac{1}{n}\Phi^\top\Phi$ is $v_{min}$,  and as stated above, $\theta \in K^{\perp}$, we have 
\begin{align*}
   \Vert \hat{\zeta} \Vert_{\sigma}^{2}=\frac{1}{n} \theta^{\top}\Phi^{\top}\Phi\theta \geq v_{min} \theta^{\top}\theta=v_{min} \Vert \theta \Vert_{2}^{2}.
\end{align*}
Thus $ \Vert \theta \Vert_{2} \leq \frac{\Vert \hat{\zeta} \Vert_{\sigma}}{\sqrt{v_{min}}}$ (positivity derived from definition). Then Equation (\ref{lemma_markovdesign_1}) adapts to
\begin{equation}
\begin{aligned}
\label{lemma_markovdesign_3}
   \Vert \hat{\zeta} \Vert_{\sigma} \leq \frac{1}{n\sqrt{v_{min}}} \left [\sum_{i=1}^{d}(\sum_{k=1}^{n}\zeta_k\Phi_{i}(X_k))^2 \right ]^{\frac{1}{2}}\end{aligned}.
\end{equation}
Then from Definition 1, for $i=1,\cdots,d$, we obtain
\begin{align*}
   \mathbb{E}[\zeta_k\Phi_i(X_k)|X_1,\ldots,X_k]=\Phi_i(X_k)\mathbb{E}[\zeta_k|X_1,\ldots,X_k]=0.
\end{align*}
The result means the term $\zeta_k\Phi_i(X_k)$ is a martingale difference sequence which can adapt to the filtration generated by $\{X_1,\ldots,X_k\}$. Thus by a refined version of Azuma's inequality from Theorem 3.13 in \cite{mcdiarmid1998concentration}, with probability $1-\delta$ we obtain
\begin{align*}
   \delta \leq 2\exp \left (-\frac{(\sum_{k=1}^{n}\zeta_k\Phi_{i}(X_k))^2}{2nC^2L^4} \right ),
\end{align*}
where $\zeta_k$ is bounded by $C$ according to Definition 1. And by definition of feature function in (\ref{feature_func}) we have $\vert \Phi_{i}(X_k) \vert \leq L^2$, thus for any $k$ and $i$
\begin{equation}
\label{lemma_markovdesign_4}    
\begin{aligned}
    \left \vert \sum_{k=1}^{n}\zeta_k\Phi_{i}(X_k) 
 \right \vert \leq CL^2\sqrt{2n\log (2/\delta)}.
\end{aligned}
\end{equation}
Combining Equations (\ref{lemma_markovdesign_3}) and (\ref{lemma_markovdesign_4}), it can be derived with probability $1-\delta$
\begin{align*}
   \Vert \hat{\zeta} \Vert_{\sigma} \leq CL^2\sqrt{\frac{2d\log(2/ \delta)}{nv_{min}}}.
\end{align*}
$\qquad \qquad \qquad \qquad \qquad \qquad \qquad \qquad \qquad \qquad  \qquad \quad \ \ \ \qedsymbol$

Next we will analyze bound of the term $\Vert \hat{\Pi}\hat{T}Q - \hat{\Pi}Q \Vert_{\sigma}$, which is induced by directly applying Lemma 2 with $\tilde{Z}=\hat{T}Q$ and $Z=Q$. 
We replace function $f$ used in Lemma 2 with action value function $q^{\beta}$, and replace $\zeta_k$ with temporal difference $ r(x_k,a_k,b_k)+\gamma \max_{a^{\prime}}\mathbb{E}_{\beta}q(x_{k+1}, a^{\prime},\beta(x_{k+1}))- q^{\beta}(x_k,a_k,b_k)$.Then we have for $1\leq k \leq n$,
\begin{align*}
    \zeta_k &= \gamma [\max_{a^{\prime}}\mathbb{E}_{\beta}q(x_{k+1}, a^{\prime},\beta(x_{k+1})) \\ & \quad - p(x^{\prime}|x_k,a_k,b_k) \max_{a^{\prime}}\mathbb{E}_{\beta}q(x^{\prime}, a^{\prime},\beta(x^{\prime}))] \\
    &\leq \gamma \max_{a^{\prime}}\mathbb{E}_{\beta}q(x_{k+1}, a^{\prime},\beta(x_{k+1}))
    \\&\leq
    \gamma q_{max}.
\end{align*}
Thus we obtain the inequality (\ref{lemma_markovdesign_4}) in Lemma 4
\begin{align*}
    \left \vert \sum_{k=1}^{n}\zeta_k\Phi_{i}(X_k) 
    \right \vert \leq \frac{\gamma L^2 q_{max}}{1-\gamma}\sqrt{2n\log (2/\delta)}.
\end{align*}
Then with probability $1-\delta$, we deduce
\begin{equation}
\label{est}
    \begin{aligned}
    \Vert \hat{\Pi}\hat{T}Q - \hat{\Pi}Q \Vert_{\sigma} \leq \frac{\gamma L^2 q_{max}}{1-\gamma} \sqrt{\frac{2d\log (2/ \delta)}{nv_{min}}}.
\end{aligned}
\end{equation}

We will discuss the value of $q_{max}$ in the next subsection such that the bound in (\ref{est}) is actually well-defined.

\subsection{Sampling error for approximate value function}
\label{error_3}
This subsection is devoted to the analysis of bound of term $\Vert \hat{Q}-\hat{Q}_s \Vert_{\sigma}$. With Assumption 1, we ensure the concentration property of transition probability. Then we need to derive the concentration property of reward function by which the bound can be deduced.

From equation (\ref{reward}) we known that the empirical reward
\begin{align*}
    \hat{r}(x_k,a_k,b_k)=\rho(x_{k-1},a_{k-1},b_{k-1})\Delta t_k,
\end{align*}
where $\rho(x_{k-1},a_{k-1},b_{k-1})$ is determined by equation (\ref{Reward}), empirical interval time $\Delta t_k$ follows a distribution related to arrival rate $\lambda$ and service rate $\mu$. 

Considering the three components of $\rho(x_{k-1},a_{k-1},b_{k-1})$, the first term $\Vert x_{k-1} \Vert_{1}$ is explicitly bounded by sum of buffer size across all servers, which is $mL$. The second and third term refer to action costs, which is also bounded and comes to zero if corresponding action is not taken. Thus $\rho_{max}=mL+c_b$.

We discuss the case for any $k$, thus we denote $\Delta t_k$ as $\Delta t$ for simplicity. We then denote the interval time until next arrival as $\Delta t^1$ with its distribution $f_a(\Delta t^1)$, the interval time until next service as $\Delta t^2$ with its distribution $f_s(\Delta t^2)$. The number of current activated servers (i.e. with job in queue) is defined as
\begin{align*}
    k_x:=k(x)=\sum_{i=0}^{m} \mathbb{I} \{x_i \neq 0 \}.
\end{align*}
It is known from property of parallel queuing system that
\begin{align*}
    f_a(\Delta t^1) &= \lambda \exp{(-\lambda \Delta t^1)}\\
    f_s(\Delta t^2) &= k_x\mu \exp{(-k_x\mu \Delta t^2)}.
\end{align*}
Since $\Delta t=\min \{ \Delta t^1, \Delta t^2 \}$, we can derived the distribution of $\Delta t$ as $f(\Delta t)$ 
\begin{align*}
    f(\Delta t)=(\lambda + k_x\mu) \exp (-(\lambda+k_x\mu)\Delta t),
\end{align*}
with expectation of $\Delta t$ as $\mathbb{E}[\Delta t]=\frac{1}{\lambda+k_x\mu}$.
Since the maximum of value function is known to be bounded with $\frac{r_{max}}{1-\gamma}$ \cite{chen2019information},we have for $q_{max}$ in (\ref{est})
\begin{align*}
    q_{max} = \frac{r_{max}}{1-\gamma}= \frac{\rho_{max}\mathbb{E}[\Delta t]}{1-\gamma} =\frac{mL+c_b}{(\lambda+k_x\mu)(1-\gamma)} \leq \frac{mL+c_b}{\lambda(1-\gamma)},
\end{align*}
where $r_{max}$ is the maximum true reward.

Next we analyze the difference between true reward and empirical reward.
Denote the bound of absolute difference between observe internal time and expected internal time (i.e.$ \frac{1}{\lambda+k_x\mu}$) as $h$, i.e.
\begin{align*}
    \left \vert \Delta\hat{t}-\mathbb{E}[\Delta t] \right \vert \leq h.
\end{align*}
We argue that with high probability(at least $1-e^{-3}$), $h\leq \frac{1}{\lambda}$.
When $h\geq \frac{1}{\lambda} \geq \frac{1}{\lambda+k_x\mu}$, the probability
\begin{align*}
    \delta &= P(\left \vert \Delta\hat{t}-\mathbb{E}[\Delta t]\right \vert \geq h) = P( \Delta\hat{t}(X) \geq \mathbb{E}[\Delta t] + h)\\
    &= \exp{\left (-(\lambda+k_x\mu)(\frac{1}{\lambda+k_x\mu}+h) \right )}\\
    &\leq \exp{\left (-1-\frac{\lambda+k_x\mu}{\lambda} \right)} \leq  \exp{\left (-2-\frac{b\mu}{\lambda} \right)} \leq e^{-3} ,
\end{align*}
where $\lambda < b\mu$ is a necessary condition for the stability of queuing system. Thus we have for any $(x,a,b)$
\begin{align}
\label{rewardd}
    \left \vert \hat{r}(x,a,b)-r(x,a,b) \right \vert \leq \left \vert \rho (x,a,b) h \right \vert \leq \frac{mL+c_b}{\lambda} .
\end{align}

\noindent{\bf Lemma 3 (Sampling error)}$\mathbf{.}$ \textit{Under Assumption 1, with probability at least $1-\delta$, the absolute difference between the empirical Bellman operator and the actual operator for  any (x,a,b) can be bounded by:}
\begin{align*}
   &\vert \hat{T}\hat{q} (x,a,b) - T \hat{q} (x,a,b) \vert 
   \\& \leq \frac{mL+c_b}{\lambda}\left (1+\frac{\gamma}{1-\gamma}C_P \sqrt{\log(1 / \delta)}w(x,a,b) \right).
\end{align*}

\noindent\textit{Proof.}
By equation (\ref{rewardd}) and the triangle inequality, we have
\begin{align*}
   &\vert \hat{T}\hat{q} (x,a,b) - T \hat{q} (x,a,b) \vert\\ &= \vert \hat{r}(x,a,b)-r(x,a,b)  + \gamma \sum_{x^{\prime}} (\hat{p}(x^{\prime}|x,a,b)\\ &\quad-p(x^{\prime}|x,a,b) )  \mathbb{E}_{\beta(b^{\prime}|x^{\prime})}[\max_{a^{\prime}}\hat{q}(x^{\prime},a^{\prime},b^{\prime})]
   \\ &\leq \vert \hat{r}(x,a,b)-r(x,a,b) \vert + \gamma \vert\sum_{x^{\prime}} (\hat{p}(x^{\prime}|x,a,b)\\ &\quad-p(x^{\prime}|x,a,b) \vert \mathbb{E}_{\beta(b^{\prime}|x^{\prime})}[\max_{a^{\prime}} \hat{q}(x^{\prime},a^{\prime},b^{\prime})] \\
   &\leq \frac{mL+c_b}{\lambda} + \gamma C_P \sqrt{\log(1 / \delta)}w(x,a,b)
   \\& \ \quad\mathbb{E}_{\beta(b^{\prime}|x^{\prime})}[\max_{a^{\prime}} \hat{q}(x^{\prime},a^{\prime},b^{\prime})]
   \\ & \leq \frac{mL+c_b}{\lambda} + \gamma C_P \sqrt{\log(1 / \delta)}w(x,a,b)q_{max} \\
   &=\frac{mL+c_b}{\lambda}\left (1+\frac{\gamma}{1-\gamma}C_P \sqrt{\log(1 / \delta)}w(x,a,b) \right).
\end{align*}
$\qquad \qquad \qquad \qquad \qquad \qquad \qquad \qquad \qquad \qquad  \qquad \quad \ \ \ \qedsymbol$

Based on the above results, we can prove Theorem 1.
\noindent\textit{Proof of Theorem 1.} 
It follows by solving equation for $\Vert Q-\hat{Q}_s \Vert_{n}$ and replacing term $\Vert \hat{\Pi}\hat{T}Q - \hat{\Pi}Q \Vert_{n}$ that
\begin{align*}
    \Vert Q-\hat{Q}_s \Vert_{\sigma} &\leq \frac{1}{\sqrt{1-\gamma^2}}\Vert Q-\hat{\Pi}Q \Vert_{\sigma}\\& \quad + \frac{\gamma L^2 (mL+c_b)}{\lambda(1-\gamma)^3}\sqrt{\frac{2d\log (2/ \delta)}{nv_{min}}}.
\end{align*}
And it can be deduced by Lemma 3 that
\begin{align*}  
  \hat{Q}_s&=\hat{\Pi}\hat{T}\Phi \theta^*  \\& \leq \hat{\Pi} T\Phi \theta^* + \hat{\Pi}\frac{mL+c_b}{\lambda}\left (\mathbf{1}_n+\frac{\gamma}{1-\gamma}C_P \sqrt{\log(1 / \delta)}W_n \right),
\end{align*}
where $\hat{\Pi} T\Phi \theta^*$ is $\hat{Q}$, $\mathbf{1}_n =[1,\ldots,1]^{\top} \in \mathbb{R}^n$. With again the triangle inequality, we have
\begin{align*}
    \Vert Q- &\hat{Q} \Vert_{\sigma}   \leq \underbrace{\frac{1}{\sqrt{1-\gamma^2}} \Vert Q-\hat{\Pi}Q \Vert_{\sigma} }_{e_p}\\&  + \underbrace{ \frac{\gamma L^2 (mL+c_b)}{\lambda(1-\gamma)^3}\sqrt{\frac{2d\log (2/ \delta)}{nv_{min}}}}_{e_{st}} \\&  + \underbrace{
    \frac{1}{\sqrt{n}} \left \Vert \hat{\Pi}\frac{mL+c_b}{\lambda}\left (\mathbf{1}_n+\frac{\gamma}{1-\gamma}C_P \sqrt{\log(1 / \delta)}W_n \right) \right \Vert_2}_{e_{sa}} .
\end{align*}
$\qquad \qquad \qquad \qquad \qquad \qquad \qquad \qquad \qquad \qquad  \qquad \quad \ \ \ \qedsymbol$

The first term  $e_p$ derives from projection error, which is unavoidable due to the finite approximate ability of feature functions. The second term $e_{st}$ is actually derived from combination of  sampling error and TD error. The third term $e_{sa}$ derived from the sampling error w.r.t approximate value function. They give hints on tuning system parameters(e.g.$n$,$m$,$d$,$c_b$,$\lambda$,$\gamma$,$L$) to decrease the prediction error.

\section{Policy Iteration}
\label{sec_pl}
In this section, we generalize the algorithm to control setting  by incorporating a minimax problem. Then, we derive the the Minimax LSPI algorithm and present its theoretical guarantee. 
\subsection{Minimax LSPI}
 Since we have obtained the policy evaluation step of Minimax LSPI in Section \ref{sec_pe}, we just need to define the policy improvement step. The optimal policy $\beta^*(x)$ with current $Q$ value is defined by:
\begin{equation}
\begin{aligned}
\label{linear_programme}
    \beta^*(x)=\arg \min_{\beta }\max_{a}\sum_{b \in B}\beta(b|x)q(x,a,b).
\end{aligned}
\end{equation}

For (\ref{linear_programme}), the identical formulation is a linear programming, where $p(b)$ specifies the action distribution of defender policy $\beta(b|x) = p(b)$ at a given state $x$. When the action set is finite, the optimum objective $c=\min_{\beta}\max_{a}\sum_{b\in B}\beta(b|x)q(x,a,b)$ is well-defined 
 \cite{zhu2020online}.
\begin{equation*}
    \left\{
        \begin{alignedat}{2}
            \min & \quad \ c \\
            s.t  &\quad \sum_{b} p(b)q(x,a,b) \leq c \qquad  \forall a\in A\\
             &\quad \ p(b) \geq 0  \qquad \forall b\in B\\
             &\quad \sum_{b} p(b)=1
        \end{alignedat}
    \right.
\end{equation*}


The pseudo code of Minimax LSPI algorithm is given.The reasons why we use updated $\mathcal{D}$ instead of a fixed datasets like \cite{lagoudakis2003least} can be listed as follows. The first one is that it's difficult to find a representative dataset for a specific system in the security domain. The second one is that a strategic attacker also learn, thus rendering the early samples useless. The third one is that additional error stems from the distributional shift between the behavior policy and the learned policy 
 \cite{kumar2020conservative}.
\begin{algorithm}[htb] 
\caption{Minimax LSPI algorithm}  
\begin{algorithmic}[1] 
\REQUIRE ~~\\ 
Initial $\theta$, initial state distribution $d$, sample memory $\mathcal{D}=\emptyset$, update steps $n$, iterations $k=0$;\\
\STATE Initialize state at $t=0$ with $x_0 \sim d$ 
    \FOR{$t=0,1,\cdots$}
        \STATE choose defender action according to $\epsilon$-greedy policy
        \begin{align*}
            u_t\sim
            \begin{cases}
                \beta_k & \ 1-\epsilon_t \ \rm{probability}\\
                random \ b\in B & \qquad \epsilon_t \ \rm{probability}
            \end{cases}
        \end{align*}
        \STATE receive $r_{t+1}$ and observe $x_{t+1}$ via action $u_t$ and opponent's action $a_t$
        \STATE calculate target value \\$y_t=r_{t+1}+\gamma\max_{a^{\prime}}\sum_{b^{\prime}} \beta_k(b^{\prime}|x_{t+1}) \mathnormal{\phi}(x_{t+1},a^{\prime} b^{\prime})\theta_k$
        \STATE store $(x_t, a_t, b_t, r_{t+1}, x_{t+1}, y_t)$ in $\mathcal{D}$
        \IF{($t$ mod $n$)==0}
            \STATE Define the loss $LS(\theta_k)=\sum_{t=1}^{n}(y_t-\mathnormal{\phi}(x_t,a_t,b_t)\theta_k)$
            \STATE update parameters with $\theta_{k+1} \gets \frac{\partial LS(\theta_k)}{\partial \theta_k}$
            \IF{$\Vert \theta_{k+1}-\theta_k \Vert < \epsilon$}
            \RETURN $\theta_k$
            \ENDIF
            \STATE $k=k+1,\mathcal{D}=\emptyset$
            \STATE use linear programming to update $\beta_k$\\
            $\beta(\cdot |x)=\arg\min_{\beta} \max_{a} \sum_{b} \beta_k(b|x) \mathnormal{\phi}(x,a,b)\theta_k$ 
        \ENDIF
    \ENDFOR
 \end{algorithmic}
\end{algorithm}

\subsection{Discussion}
Minimax LSPI is actually an extension of the LSPI algorithm that not only satisfies the previously discussed performance bound, but also possesses same properties as LSPI. Hence, Minimax LSPI is also an approximate policy-iteration algorithm. Its fundamentally soundness has already been verified by \cite{bertsekas1996neuro}. Furthermore, the error in policy improvement is eliminated since it implicitly represents a policy as Q functions and chooses essentially greedy (mixed) action, at cost of extra optimization. 

Theorem 1 established the bound of prediction error for sample set without requirement of specific policy and stationary distribution. Then 
let $ \{\beta_j \}_{j=1}^{n}$ be the sequence of policies generated by Minimax LSPI. $ \{\hat{Q}^{\beta_j} \}_{j=1}^{n}$ are corresponding approximate value functions. If there exists a bound $\epsilon$ for prediction error s.t for any $\beta_j$
\begin{align*}
    \Vert Q^{\beta_j}-\hat{Q}^{\beta_j} & \Vert_{\infty} \leq \epsilon;
\end{align*}
Then a near-optimal policy can be learned with:
\begin{align*}
    \limsup_{j \xrightarrow{} \infty} \Vert Q^*-\hat{Q}^{\beta_j} & \Vert_{\infty} \leq \frac{2\gamma\epsilon}{(1-\gamma)^2}.
\end{align*}

The convergence guarantee is essentially adapted from \cite[Theorem 7.1]{lagoudakis2003least}, which also shows that the learned policy can converge to the near equilibrium strategy of the Markov game.

From Theorem 1, we can deduce that the bound $\epsilon$ is determined also by projection error and sampling error. While the sampling error can be reduced or even eliminated by tuning system parameters. Then an appropriate choice of feature functions can provide theoretical guarantees for quality of learned policy by Minimax LSPI.

%% file: Texts/conclusion.tex
\section{Conclusion and future work}
\label{sec_con}
In this paper, we generalize LSPI to Markov game with incomplete information and propose Minimax LSPI algorithm, incorporating a minimax policy improvement implementation. Since if the prediction error of LSPI is bounded across state-action space, we can guarantee policy improvement and its convergence to an at least near-optimal policy. Our work focus on the bound of prediction error of LSPI in multi-agent settings. We decompose prediction error to projection error and sampling error, then we analyze the bound of each separately. Our main results give a concrete bound of the prediction error over sample sets of Minimax LSPI. The result also provides insights for tuning system parameters. The ongoing work will be the generalization of error bound from $\sigma$ norm to $l_\infty$ norm.




%% file: main.bbl
\begin{thebibliography}{10}
\providecommand{\url}[1]{#1}
\csname url@samestyle\endcsname
\providecommand{\newblock}{\relax}
\providecommand{\bibinfo}[2]{#2}
\providecommand{\BIBentrySTDinterwordspacing}{\spaceskip=0pt\relax}
\providecommand{\BIBentryALTinterwordstretchfactor}{4}
\providecommand{\BIBentryALTinterwordspacing}{\spaceskip=\fontdimen2\font plus
\BIBentryALTinterwordstretchfactor\fontdimen3\font minus \fontdimen4\font\relax}
\providecommand{\BIBforeignlanguage}[2]{{%
\expandafter\ifx\csname l@#1\endcsname\relax
\typeout{** WARNING: IEEEtran.bst: No hyphenation pattern has been}%
\typeout{** loaded for the language `#1'. Using the pattern for}%
\typeout{** the default language instead.}%
\else
\language=\csname l@#1\endcsname
\fi
#2}}
\providecommand{\BIBdecl}{\relax}
\BIBdecl

\bibitem{jin2018stability}
L.~Jin and S.~Amin, ``{Stability of Fluid Queueing Systems with Parallel Servers and Stochastic Capacities},'' \emph{IEEE Transactions on Automatic Control}, vol.~63, no.~11, pp. 3948--3955, 2018.

\bibitem{fraile2018trustworthy}
F.~Fraile, T.~Tagawa, R.~Poler, and A.~Ortiz, ``{Trustworthy Industrial IoT Gateways for Interoperability Platforms and Ecosystems},'' \emph{IEEE Internet of Things Journal}, vol.~5, no.~6, pp. 4506--4514, 2018.

\bibitem{laszka2019detection}
A.~Laszka, W.~Abbas, Y.~Vorobeychik, and X.~Koutsoukos, ``{Detection and Mitigation of Attacks on Transportation Networks as a Multi-stage Security Game},'' \emph{Computers \& Security}, vol.~87, p. 101576, 2019.

\bibitem{feng2022cybersecurity}
Y.~Feng, S.~E. Huang, W.~Wong, Q.~A. Chen, Z.~M. Mao, and H.~X. Liu, ``{On the Cybersecurity of Traffic Signal Control System with Connected Vehicles},'' \emph{IEEE Transactions on Intelligent Transportation Systems}, vol.~23, no.~9, pp. 16\,267--16\,279, 2022.

\bibitem{manshaei2013game}
M.~H. Manshaei, Q.~Zhu, T.~Alpcan, T.~Bac{\c{s}}ar, and J.-P. Hubaux, ``{Game Theory Meets Network Security and Privacy},'' \emph{ACM Computing Surveys (CSUR)}, vol.~45, no.~3, pp. 1--39, 2013.

\bibitem{xie2024cost}
Q.~Xie, J.~Wang, and L.~Jin, ``{Cost-aware Defense for Parallel Server Systems against Reliability and Security Failures},'' \emph{Automatica}, vol. 160, p. 111467, 2024.

\bibitem{xu2016playing}
H.~Xu, L.~Tran-Thanh, and N.~Jennings, ``{Playing Repeated Security Games with no Prior Knowledge},'' in \emph{AAMAS'16: Proceedings of the 2016 International Conference on Autonomous Agents \& Multiagent Systems}.\hskip 1em plus 0.5em minus 0.4em\relax ACM Press, 2016, pp. 104--112.

\bibitem{zhang2023security}
H.~Zhang, Y.~Mi, Y.~Fu, X.~Liu, Y.~Zhang, J.~Wang, and J.~Tan, ``{Security Defense Decision Method Based on Potential Differential Game for Complex Networks},'' \emph{Computers \& Security}, vol. 129, p. 103187, 2023.

\bibitem{sutton2018reinforcement}
R.~S. Sutton and A.~G. Barto, \emph{{Reinforcement Learning: An Introduction}}.\hskip 1em plus 0.5em minus 0.4em\relax MIT press, 2018.

\bibitem{shapley1950basic}
L.~S. Shapley and R.~Snow, ``{Basic Solutions of Discrete Games},'' \emph{Contributions to the Theory of Games}, vol.~1, no.~24, pp. 27--27, 1950.

\bibitem{fudenberg2016whither}
D.~Fudenberg and D.~K. Levine, ``{Whither Game Theory? Towards a Theory of Learning in Games},'' \emph{Journal of Economic Perspectives}, vol.~30, no.~4, pp. 151--170, 2016.

\bibitem{brown1951iterative}
G.~W. Brown, ``{Iterative Solution of Games by Fictitious Play},'' \emph{Act. Anal. Prod Allocation}, vol.~13, no.~1, p. 374, 1951.

\bibitem{littman1994markov}
M.~L. Littman, ``{Markov Games as a Framework for Multi-agent Reinforcement Learning},'' in \emph{Machine learning proceedings 1994}.\hskip 1em plus 0.5em minus 0.4em\relax Elsevier, 1994, pp. 157--163.

\bibitem{bowling2002multiagent}
M.~Bowling and M.~Veloso, ``{Multiagent Learning Using a Variable Learning Rate},'' \emph{Artificial intelligence}, vol. 136, no.~2, pp. 215--250, 2002.

\bibitem{zhang2020model}
K.~Zhang, S.~Kakade, T.~Basar, and L.~Yang, ``{Model-based Multi-agent RL in Zero-sum Markov Games with Near-optimal Sample Complexity},'' \emph{Advances in Neural Information Processing Systems}, vol.~33, pp. 1166--1178, 2020.

\bibitem{perolat2016softened}
J.~P{\'e}rolat, B.~Piot, M.~Geist, B.~Scherrer, and O.~Pietquin, ``{Softened Approximate Policy Iteration for Markov Games},'' in \emph{International Conference on Machine Learning}.\hskip 1em plus 0.5em minus 0.4em\relax PMLR, 2016, pp. 1860--1868.

\bibitem{mao2023provably}
W.~Mao and T.~Ba{\c{s}}ar, ``{Provably Efficient Reinforcement Learning in Decentralized General-sum Markov Games},'' \emph{Dynamic Games and Applications}, vol.~13, no.~1, pp. 165--186, 2023.

\bibitem{sayin2021decentralized}
M.~Sayin, K.~Zhang, D.~Leslie, T.~Basar, and A.~Ozdaglar, ``{Decentralized Q-learning in Zero-sum Markov Games},'' \emph{Advances in Neural Information Processing Systems}, vol.~34, pp. 18\,320--18\,334, 2021.

\bibitem{lagoudakis2003least}
M.~G. Lagoudakis and R.~Parr, ``{Least-squares Policy Iteration},'' \emph{The Journal of Machine Learning Research}, vol.~4, pp. 1107--1149, 2003.

\bibitem{lazaric2012finite}
A.~Lazaric, M.~Ghavamzadeh, and R.~Munos, ``{Finite-sample Analysis of Least-squares Policy Iteration},'' \emph{Journal of Machine Learning Research}, vol.~13, pp. 3041--3074, 2012.

\bibitem{osband2017posterior}
I.~Osband and B.~Van~Roy, ``{Why is Posterior Sampling Better than Optimism for Reinforcement Learning?}'' in \emph{International conference on machine learning}.\hskip 1em plus 0.5em minus 0.4em\relax PMLR, 2017, pp. 2701--2710.

\bibitem{gallager2013stochastic}
R.~G. Gallager, \emph{{Stochastic Processes: Theory for Applications}}.\hskip 1em plus 0.5em minus 0.4em\relax Cambridge University Press, 2013.

\bibitem{shapley1953stochastic}
L.~S. Shapley, ``{Stochastic Games},'' \emph{Proceedings of the national academy of sciences}, vol.~39, no.~10, pp. 1095--1100, 1953.

\bibitem{zhu2020online}
Y.~Zhu and D.~Zhao, ``{Online Minimax Q Network Learning for Two-player Zero-sum Markov Games},'' \emph{IEEE Transactions on Neural Networks and Learning Systems}, vol.~33, no.~3, pp. 1228--1241, 2020.

\bibitem{lazaric2010finite}
A.~Lazaric, M.~Ghavamzadeh, and R.~Munos, ``{Finite-sample Analysis of LSTD},'' in \emph{ICML-27th International Conference on Machine Learning}, 2010, pp. 615--622.

\bibitem{kumar2020conservative}
A.~Kumar, A.~Zhou, G.~Tucker, and S.~Levine, ``{Conservative Q-learning for Offline Reinforcement Learning},'' \emph{Advances in Neural Information Processing Systems}, vol.~33, pp. 1179--1191, 2020.

\bibitem{auer2008near}
P.~Auer, T.~Jaksch, and R.~Ortner, ``{Near-optimal Regret Bounds for Reinforcement Learning},'' \emph{Advances in neural information processing systems}, vol.~21, 2008.

\bibitem{van1998learning}
B.~Van~Roy, ``{Learning and Value Function Approximation in Complex Decision Processes},'' Ph.D. dissertation, Massachusetts Institute of Technology, 1998.

\bibitem{tsitsiklis1996analysis}
J.~Tsitsiklis and B.~Van~Roy, ``{Analysis of Temporal-diffference Learning with Function Approximation},'' \emph{Advances in neural information processing systems}, vol.~9, 1996.

\bibitem{mcdiarmid1998concentration}
C.~McDiarmid, ``{Concentration, Probabilistic Methods for Algorithmic Discrete Mathematics, 195--248},'' \emph{Algorithms Combin}, vol.~16, 1998.

\bibitem{chen2019information}
J.~Chen and N.~Jiang, ``{Information-theoretic Considerations in Batch Reinforcement Learning},'' in \emph{International Conference on Machine Learning}.\hskip 1em plus 0.5em minus 0.4em\relax PMLR, 2019, pp. 1042--1051.

\bibitem{bertsekas1996neuro}
D.~Bertsekas and J.~N. Tsitsiklis, \emph{{Neuro-dynamic Programming}}.\hskip 1em plus 0.5em minus 0.4em\relax Athena Scientific, 1996.

\end{thebibliography}
